\documentclass[11pt, a4paper, twoside]{article}
\pdfoutput=1
\usepackage[square,numbers]{natbib}
\usepackage[english]{babel}
\usepackage[utf8]{inputenc}
\usepackage[font=small,labelfont=bf]{caption}
\usepackage{latexsym}
\usepackage{amsfonts}
\usepackage{amsmath}
\usepackage{amsthm}
\usepackage{amssymb}
\usepackage{fancyhdr}
\usepackage{esdiff}
\usepackage{empheq}
\usepackage{graphicx}
\usepackage{mathrsfs}
\usepackage{wrapfig}
\usepackage[all]{xy}
\usepackage{comment}
\usepackage{enumerate}
\usepackage[babel]{csquotes}
\usepackage{geometry}
\usepackage[Lenny]{fncychap}
\usepackage{xcolor}
\usepackage{epstopdf}
\usepackage[linktoc=page,colorlinks,
linkcolor=blue1,citecolor=green,urlcolor=blue1,runcolor=red1,linkbordercolor={0 0 1},urlbordercolor={1 0 0},citebordercolor={0 1 0},runbordercolor={1 1 1}]{hyperref}
\usepackage{tikz}
\usepackage{tcolorbox}
\usetikzlibrary{decorations.markings}
\usetikzlibrary{calc}
\usepackage{marvosym}

\DeclareFontFamily{U}{mathx}{\hyphenchar\font45}
\DeclareFontShape{U}{mathx}{m}{n}{
      <5> <6> <7> <8> <9> <10>
      <10.95> <12> <14.4> <17.28> <20.74> <24.88>
      mathx10
      }{}
\DeclareSymbolFont{mathx}{U}{mathx}{m}{n}
\DeclareFontSubstitution{U}{mathx}{m}{n}
\DeclareMathAccent{\widecheck}{0}{mathx}{"71}

\makeatletter
\let\save@mathaccent\mathaccent
\newcommand*\if@single[3]{%
  \setbox0\hbox{${\mathaccent"0362{#1}}^H$}%
  \setbox2\hbox{${\mathaccent"0362{\kern0pt#1}}^H$}%
  \ifdim\ht0=\ht2 #3\else #2\fi
  }
\newcommand*\rel@kern[1]{\kern#1\dimexpr\macc@kerna}
\newcommand*\widebar[1]{\@ifnextchar^{{\wide@bar{#1}{0}}}{\wide@bar{#1}{1}}}
\newcommand*\wide@bar[2]{\if@single{#1}{\wide@bar@{#1}{#2}{1}}{\wide@bar@{#1}{#2}{2}}}
\newcommand*\wide@bar@[3]{%
  \begingroupand vague
  \def\mathaccent##1##2{%
    \let\mathaccent\save@mathaccent
    \if#32 \let\macc@nucleus\first@char \fi
    \setbox\z@\hbox{$\macc@style{\macc@nucleus}_{}$}%
    \setbox\tw@\hbox{$\macc@style{\macc@nucleus}{}_{}$}%
    \dimen@\wd\tw@
    \advance\dimen@-\wd\z@
    \divide\dimen@ 3
    \@tempdima\wd\tw@
    \advance\@tempdima-\scriptspace
    \divide\@tempdima 10
    \advance\dimen@-\@tempdima
    \ifdim\dimen@>\z@ \dimen@0pt\fi
    \rel@kern{0.6}\kern-\dimen@
    \if#31
      \overline{\rel@kern{-0.6}\kern\dimen@\macc@nucleus\rel@kern{0.4}\kern\dimen@}%
      \advance\dimen@0.4\dimexpr\macc@kerna
      \let\final@kern#2%
      \ifdim\dimen@<\z@ \let\final@kern1\fi
      \if\final@kern1 \kern-\dimen@\fi
    \else
      \overline{\rel@kern{-0.6}\kern\dimen@#1}%
    \fi
  }%
  \macc@depth\@ne
  \let\math@bgroup\@empty \let\math@egroup\macc@set@skewchar
  \mathsurround\z@ \frozen@everymath{\mathgroup\macc@group\relax}%
  \macc@set@skewchar\relax
  \let\mathaccentV\macc@nested@a
  \if#31
    \macc@nested@a\relax111{#1}%
  \else
    \def\gobble@till@marker##1\endmarker{}%
    \futurelet\first@char\gobble@till@marker#1\endmarker
    \ifcat\noexpand\first@char A\else
      \def\first@char{}%
    \fi
    \macc@nested@a\relax111{\first@char}%
  \fi
  \endgroup
}

\fancypagestyle{subsection}{
\fancyhf{}

\fancyhead[
RO]
{\nouppercase{\small\leftmark}}
\fancyhead[
LE]
{\nouppercase{\small\rightmark}}
\fancyfoot[RO,LE]{\thepage}}

\fancypagestyle{section}{
\fancyhf{}

\fancyhead[LO,RE]{}
\fancyhead[RO,LE]{\nouppercase{\small\leftmark}}
\fancyfoot[RO,LE]{\thepage}}

\fancypagestyle{plain}{
\fancyhf{}

\fancyhead[LO,RE]{}
\fancyhead[RO,LE]{}
\fancyfoot[RO,LE]{\thepage}}

\def\beq{\begin{equation}\begin{aligned}}
\def\eeq{\end{aligned}\end{equation}}

\definecolor{blue}{rgb}{0,0,0.6}
\definecolor{blue1}{rgb}{0,0,1.5}
\definecolor{red}{rgb}{0.85,0,0}
\definecolor{red1}{rgb}{1,0,0}
\definecolor{green}{rgb}{0,0.7,0}

\theoremstyle{definition}
\newtheorem{definition}{Definition}[section]
\theoremstyle{definition}

\theoremstyle{definition}

\theoremstyle{definition}

\theoremstyle{definition}

\theoremstyle{definition}

\theoremstyle{definition}

\theoremstyle{definition}
\newtheorem{remark}[definition]{Remark}

\def\F{\mathcal F}
\def\x{\mathbf x}
\def\b{\mathbf b}
\def\g{\mathfrak g}
\def\c{\mathbf c}
\def\superPhi{\mathbf \Phi}
\def\superXi{\mathbf \Xi}
\def\L{{\cal L}}
\def\d{{\mathrm{d}}}
\def\ev{{\mathrm{ev}}}
\def\Ev{{\mathrm{Ev}}}
\def\M{{\cal M}}
\def\Map{{\mathrm{Map}}}
\def\ag{{\mathrm{ag}}}
\def\gh{{\mathrm{gh}}}
\def\EEv{{\mathbb{E}\mathrm{v}}}

\begin{document}

\title{Observables in the equivariant $A$-model}
\author{
F. Bonechi\footnote{\small INFN Sezione di Firenze, email: francesco.bonechi@fi.infn.it}, 
A.S. Cattaneo{\footnote{\small University of Zurich, email: cattaneo@math.uzh.ch}}, 
R. Iraso{\footnote{\small SISSA, email: riraso@sissa.it}},
M.Zabzine{\footnote{\small Department of Physics and Astronomy, Uppsala, maxim.zabzine@physics.uu.se}}
}

\date{\today}

\maketitle

\abstract{
We discuss observables of an equivariant extension of the $A$-model in the framework of the AKSZ construction. 
We introduce the {\it $A$-model observables}, a class of observables that are homotopically equivalent to the canonical {\it AKSZ observables} but are better behaved in the gauge fixing. 
We discuss them for two different choices of gauge fixing: the first one is conjectured to compute the correlators of the $A$-model with target the Marsden-Weinstein reduced space; in the second one we recover the topological Yang-Mills action coupled with $A$-model so that the $A$-model observables are closed under supersymmetry.
}

\pagestyle{section}
\section{Introduction}

The AKSZ method~\cite{AKSZ:geometry_of_BV} is a very elegant geometrical construction of solutions of the classical master equation (CME) in the Batalin-Vilkovisky (BV) formalism. 
It gives solutions in terms of geometrical data that are very compactly formulated in the language of graded geometry. 
The AKSZ space of fields is the space of maps from the source graded manifold $T[1]\Sigma$, where $\Sigma$ is a $d$-dimensional manifold, to the target~$\M$, which is a degree $(d-1)$ graded symplectic manifold endowed with a degree one hamiltonian vector field $D=\{\Theta,-\}$ such that $D^2=0$. 
The solution of the CME, even for classical actions whose gauge invariance is very intricated, can be obtained on the spot directly from these data, without using the tools of homological perturbation theory: see for instance~\cite{roytenberg:courant2} for an introduction to the subject and the discussion of the Courant Sigma Model. 

In the BV setting the simplest version of gauge fixing is realized by expressing the antifields as functions of the fields; once that the symplectic interpretation is taken into account and the space of fields is seen as an odd symplectic manifold, the gauge fixing is a choice of a lagrangian submanifold $\L$ of the BV space of fields~$\F$. 
Even if the BV vector field $Q_{\mathrm{BV}}$ is not in general parallel to $\L$, the gauge fixed action still has an odd symmetry obtained by projecting $Q_{\mathrm{BV}}$ to~$\L$. 
We call this odd vector field of~$\L$ the {\it residual BV symmetry}. 
This projection is not unique and depends on an additional geometrical datum, the choice of a symplectic tubular neighborhood of~$\L$, {\it i.e.} a (local) identification of~$\F$ with~$T^*[-1]\L$. 
This choice can always be done, although in a non unique way; different choices coincide \emph{on shell}, {\it i.e.} when restricted to the surface of solutions of equations of motion. 
In examples this odd symmetry of the gauge fixed action is an interesting object and so it is worth to take it into account in the full picture.  
For instance in the BV treatment of ordinary gauge theories it is the BRST differential; in the $A$-model it is the supersymmetry~\cite{bci:PSMvsA-mod}. 

A relevant aspect where one can appreciate the beauty of the AKSZ solution is the construction of observables. 
Indeed, there is a chain map from the complex of the homological vector field $D$ of the target $\M$ to the complex of $Q_{\mathrm{BV}}$ that defines the so called {\it AKSZ observables}.
Unfortunately, in general we cannot expect that after gauge fixing a BV observable is closed under the residual BV symmetry and AKSZ observables are not special in this regard. 
So in certain cases, it can be useful to introduce an equivalent set of observables that have a better behavior for the gauge fixing.

This study began in~\cite{bci:PSMvsA-mod} for the case of the $A$-model, seen as a complex  gauge fixing of the Poisson Sigma Model with non degenerate target. 
In this case, the target graded manifold is just $T[1]M$ with $M$ symplectic and $D$ the de~Rham vector field of~$M$; AKSZ observables are then defined in terms of closed forms on~$M$. 
In~\cite{bci:PSMvsA-mod} it was shown that one can define an equivalent class of observables, which we called {\it $A$-model observables}, related by an explicit homotopy to the AKSZ ones, that are closed under the residual BV symmetry fixed by the complex gauge fixing. 
The name is due to the fact that they reproduce Witten's hierarchy of observables for the $A$-model in~\cite{witten:topological}. 

In this paper we extend the analysis to an equivariant version of the Poisson Sigma Model. 
This is an AKSZ theory that was studied in~\cite{BonechiCabreraZabzine,signori,zucchini:gaugingPSM}. 
The geometrical data of the target encode a hamiltonian $G$ space, {\it i.e.} a symplectic manifold $M$ with an action of a Lie group $G$ with an equivariant momentum map~$\mu$. 
The target homological vector field encodes the Weil model for equivariant geometry. 
In~\cite{BonechiCabreraZabzine} this theory was considered as a model for the PSM with target the symplectic reduction~$\mu^{-1}(0)/G$. 
We introduce the analogue of $A$-model observables that depend on a minimal set of fields and introduce an explicit homotopy with the AKSZ observables.

We consider two different gauge fixings which are compatible with the $A$-model observables. 
The first one is relevant when the symplectic reduction of the target space is smooth; we conjecture that the theory computes the $A$-model correlator of the reduced symplectic manifold in the spirit of~\cite{BonechiCabreraZabzine}. 
In the second one, we recover for the Lie algebra sector the supersymmetric Yang Mills action and the residual BV symmetry is the supersymmetry generator.

\subsubsection*{Aknowledgements}

A. S. C. acknowledges partial support of SNF Grant No. 200020- 172498/1. This research was (partly) supported by the NCCR SwissMAP, funded by the Swiss National Science Foundation, and by the COST Action MP1405 QSPACE, supported by COST (European Cooperation in Science and Technology). 
F.B. acknowledges partial support by COST Action MP1405 QSPACE, supported by COST (European Cooperation in Science and Technology), (STSM 40249).
M. Z.  acknowledges the partial support by Vetenskapsr\r{a}det under grant \#2014-5517, by the STINT grant, and by the grant  ``Geometry and Physics"  from the Knut and Alice Wallenberg foundation.

\section{AKSZ background}\label{AKSZ_background}

In this section we review for completeness the AKSZ construction and the \emph{residual symmetry} of the gauge-fixed action in BV theories. 
See~\cite{cattaneo:supergeometry,roytenberg:courant2} for an introduction to BV in the language of graded geometry and in particular to the AKSZ construction; see~\cite{bci:PSMvsA-mod} for more details on residual BV symmetry.

A classical BV theory consists of a $(-1)$-symplectic manifold $(\F,\Omega)$ endowed with a cohomological hamiltonian vector field $Q_{\mathrm{BV}}= \{S_{\mathrm{BV}},-\}$ with degree~$1$\,, where $S_{\mathrm{BV}}$ is the BV action of the theory and $\{~,~\}$ are the Poisson brackets induced by the symplectic structure~$\Omega$.
Since $Q_{\mathrm{BV}}^2=0$ the BV-action is a solution of the classical master equation~(CME) 
\beq\label{CME}
 \{S_{\mathrm{BV}},S_{\mathrm{BV}}\}=0 ~.
\eeq
If we introduce local Darboux coordinates $\{x,x^+\}$ the bracket reads
\[
 \{F,G\} = \frac{\partial_r F}{\partial x^a}\frac{\partial_\ell G}{\partial x^+_a}-\frac{\partial_r F}{\partial x^+_a}\frac{\partial_\ell G}{\partial x^a} ~,
\]
where $\partial_r$ and $\partial_\ell$ denote the right and left derivative, respectively.%
\footnote{
 In the following, where not indicated otherwise, we will always use left derivatives.
}
The CME is expressed in these local coordinates as:
\beq\label{CME_coordinates}
 \frac{1}{2}\{S_{\mathrm{BV}},S_{\mathrm{BV}}\}= \frac{\partial_r S_{\mathrm{BV}}}{\partial x^+_a}\frac{\partial_{\ell} S_{\mathrm{BV}}}{\partial x^a}=0 ~.
\eeq

The gauge-fixing is performed  by restricting the action to a Lagrangian submanifold $\mathcal{L}\subset \F$, {\it i.e.} a submanifold on which the restriction of the symplectic form vanishes and that cannot be properly enlarged to a submanifold with this property. 
Locally we can choose Darboux coordinates $(x,x^+)$ in which $\mathcal L$ is determined by $x^+=0$. 
The gauge fixed action is just the restriction $S_\L$ to $\L$ of $S_{\mathrm{BV}}$, {\it i.e.} in these Darboux coordinates: $S_\L(x^a)=S_{\mathrm{BV}}(x^a,x^+_a=0)$.

\subsection{The residual BV symmetry}

The BV vector field $Q_{\mathrm{BV}}$ is in general not parallel to the gauge fixing lagrangian~$\L$; nevertheless it can be projected to a vector field over~$\L$ in such a way that the result is a symmetry of the gauge fixed action~$S_\L$. 
This can be done by choosing a \emph{symplectic tubular neighbourhood} of the Lagrangian, {\it i.e.} a local symplectomorphism $\F\supseteq\mathcal{U}\simeq T^*[-1]\L$ 
restricting to the identity on~$\L$\,. 
If we denote by $\iota\colon \mathcal{L} \hookrightarrow \F$ the inclusion map and with $\pi\colon\mathcal{U}\rightarrow\L$ the projection map, the residual symmetry can be then defined by:
\beq
 Q^\pi_{\mathcal{L}} := \iota^* \circ Q_{\mathrm{BV}}\circ \pi^* ~,
\eeq
where we view vector fields as operators on functions. 
More concretely, we can think of this tubular neighbourhood as an atlas of canonical coordinates $\{x,x^+\}$ adapted to~$\mathcal{L}$ ({\it i.e.} $\mathcal{L}= \{x^+=0\}$) such that the transition functions between $(x,x^+)$ and $(y,y^+)$ are $(y=y(x),y^+=(\partial x/\partial y) x^+)$ so that the projection $\pi(x,x^+)=x$ is well defined. 
For every function~$f$ on~$\mathcal{L}$ we have:
\beq
 Q^\pi_{\mathcal{L}}(f) = Q_{\mathrm{BV}}(\pi^*f) \big|_{x^+=0} = -\frac{\partial_r S_{\mathrm{BV}}}{\partial x^+_a}\bigg|_{x^+=0}\frac{\partial_{\ell} f }{\partial x^a}~.
\eeq
In particular, it follows that $Q^\pi_\L(S_{\L})=0$\,, because of the CME~\eqref{CME}.

The odd version of Weinstein's theorem on the existence of a local symplectomorphism between a neighbourhood of a Lagrangian submanifold and $T^*[-1]\mathcal{L}$ was proved in~\cite{schwarz:BV_geometry}.
It must be pointed out that such a choice is non canonical and non unique: each symplectomorphism of~$\F$ into itself which keeps~$\mathcal{L}$ fixed defines a new symplectic tubular neighbourhood. 
Nevertheless, two such vector fields coincide when restricted to the space of solutions of the equations of motion of~$S_\L$.

The residual symmetry squares to zero only \emph{on-shell},~{\it i.e.}
\beq\label{res.symm.nilp.}
 \frac{1}{2}[Q^\pi_{\mathcal{L}},Q^\pi_{\mathcal{L}}] = \sigma^{ab} \, \frac{\partial
  S_\mathcal{L}}{\partial x^b}\frac{\partial}{\partial x^a}	~,
\eeq
where $\sigma^{ab}$ is the quadratic term in the antifield expansion of the action:
\beq\label{sviluppo_azione}
 S_{\mathrm{BV}}(x,x^+) =  S_\mathcal{L}(x) -  Q^{\pi\,a}_{\mathcal{L}}(x) x^+_a + 
  \frac{1}{2}x^+_a\sigma^{ab}(x) x^+_b + O(x^{+\,3})	~.
\eeq

A BV observable by definition is a function $f$ on $\F$ that is closed under $Q_{\mathrm{BV}}$; it is clear that the restriction of $f$ to a Lagrangian submanifold $\L$ is not closed with respect to $Q_\L$; indeed we see that
\beq\label{restriction_of_observables}
 Q^\pi_\mathcal{L} (f)|_{\mathcal{L}} + V_f(S_{\mathrm{BV}})=0 ~,
\eeq
where $V_f:=\frac{\partial_r f}{\partial x^+_i}\Big|_{\mathcal{L}}\frac{\partial}{\partial x^i} \in \mathfrak{X}(\mathcal{L})$\,.
Therefore $f_\L=f|_\L$ is $Q^\pi_\mathcal{L}$-closed modulo equations of motion.

\subsection{AKSZ construction}

The AKSZ solution of the CME~\eqref{CME} is given in terms of the following data. A source graded manifold $T[1]\Sigma$ with $\Sigma$ a $d$-dimensional manifold with its canonical de~Rham vector field $\d_\Sigma$; a target graded manifold $\M$ with an exact symplectic structure $\Omega = \mathrm{d}\vartheta$ of degree $(d-1)$ and a hamiltonian vector field~$D$ of degree~$1$ with hamiltonian $\Theta$ squaring to zero. 
The degree of $\Theta$ is then fixed to $d$. 
The space of BV fields is $\F_\Sigma={\rm Map}(T[1]\Sigma,\M)$. 
If we introduce the coordinates $\{u^\alpha,\theta^\alpha\}$ of degree $(0,1)$ in $T[1]\Sigma$ and $\{x^A\}$ in $\M$, then $\F_\Sigma$ is described by the superfields
\[
 \x^A= x^A+x^A_\alpha \theta^\alpha+ \ldots ~.
\]
The evaluation map $\EEv:\F_\Sigma\times T[1]\Sigma\rightarrow\M$ is defined as
\begin{equation} \label{evaluation_map}
 \EEv(\x;u,\theta) = \x(u,\theta)~.
\end{equation}
The BV vector field is 
\begin{equation}\label{AKSZ_BV_vector field}
 Q_{\mathrm{BV}} = D'-\mathrm{d}_{\Sigma}' ~,
\end{equation}
where $D'$ and $\mathrm{d}_\Sigma'$ are the vector fields of $\F_\Sigma$ obtained by composing the maps of $\F_\Sigma$ with the target and source infinitesimal diffeomorphisms defined respectively by $D$ and $\mathrm{d}_\Sigma$. 
It is a hamiltonian vector field with hamiltonian given by
\[
 S_{\mathrm{BV}} = -\int_{T[1]\Sigma} \vartheta_A(\x)D\x^A + \int_{T[1]\Sigma} \EEv^*(\Theta) ~.
\]
By construction $S_{\mathrm{BV}}$ solves the CME~\eqref{CME}. 
Let $\omega\in C(\M)$ then we see that
\[
 (Q_{\mathrm{BV}}+\mathrm{d}_\Sigma)\EEv^*(\omega) = \EEv^*(D\omega)~,
\]
so that if $D\omega=0$ then, for each $k$-cycle $\gamma_k$ of $\Sigma$, ${\cal O}_{\omega\gamma_k} = \int_{\gamma_k}\EEv^*\omega$ is $Q_{\mathrm{BV}}$~closed. 
We say that ${\cal O}_\omega=\EEv^*\omega$ is the {\it AKSZ observable} associated to~$\omega$.

\section{$A$-model and PSM correspondence reconsidered}

The correspondence between the AKSZ observables of the PSM and the observables of the $A$-model established in~\cite{bci:PSMvsA-mod} can be better understood starting from an homotopy between maps of superspaces. 

Let $M$ be a symplectic manifold and let us denote with $\alpha=\alpha_{\mu\nu} \d x^\mu \d x^\nu$ the symplectic form. 
The Poisson Sigma Model (PSM) with non degenerate target is the AKSZ construction with a two-dimensional source manifold $\Sigma$ and target $T^*[1]M$ with hamiltonian $\alpha^{\mu\nu}b_\mu b_\nu$, where $\{x^\mu, b_\mu\}$ are the degree $(0,1)$ coordinates of $T^*[1]M$ and $\alpha^{\mu\nu}$ is the inverse of $\alpha_{\mu\nu}$. 

The space of AKSZ field is ${\cal F}_\Sigma=\mathrm{Maps}(T[1]\Sigma, T^*[1]M)$.
The symplectic form identifies it with $\mathrm{Maps}(T[1]\Sigma, T[1]M)$ and finally with $T[1](\mathrm{Maps}(T[1]\Sigma, M))\equiv T[1]{\cal M}_\Sigma$.
With this identification observables are forms on ${\cal M}_\Sigma$. 
Let the superfields $(\mathbf{x},\mathbf{b})\in{\cal F}_\Sigma$ be decomposed as
\[
 \mathbf{x}^\mu = x^\mu +\eta^{+\mu} + b^{+\mu}	~,\qquad \mathbf{b}_\mu = b_\mu + \eta_\mu + x^+_\mu ~.
\]
The de~Rham differential $\delta$ of ${\cal M}_\Sigma$ acts as $\delta\mathbf{x}^\mu=\mathbf{b}^\mu\equiv\alpha(\mathbf{x})^{\mu\nu}\mathbf{b}_\nu$. 
It can also be interpreted as the (infinitesimal) diffeomorphism obtained by composing the superfields with the (infinitesimal) diffeomorphism of the target $T[1]M$ defined by the de~Rham differential. 
The BV differential is then defined as 
\[
 Q_{\mathrm{BV}}=\delta {-}\mathrm{d}'_\Sigma~,
\] 
where $\mathrm{d}'_\Sigma$ is the vector field of $\F_\Sigma$ obtained by the action of the de~Rham differential of $\Sigma$ on the superfields. 
More geometrically, $\mathrm{d}'_\Sigma\in{\rm Vect}(\F_\Sigma)$ is the (infinitesimal) diffeomorphism of $\F_\Sigma$ obtained by composing maps with the (infinitesimal) diffeomorphism of the source defined by the de~Rham differential. 
Although ${\mathrm d}_\Sigma'$ must not be confused with ${\mathrm d}_\Sigma$ acting on $\Sigma$, they coincide on functions of the (evaluated) superfields,~{\it i.e.}
\beq
 (\mathrm{d}'_\Sigma - \mathrm{d}_\Sigma)f(\mathbf{x}(u,\theta),\mathbf{b}(u,\theta))=0 ~.
\eeq 
It is explicitly given by the following formulas:
\beq
\begin{aligned}
 &\mathrm{d}'_{\Sigma}x = 0 ~, 				\qquad &&\mathrm{d}'_{\Sigma}b = 0 ~,\\
 &\mathrm{d}'_{\Sigma}\eta^+ = \mathrm{d}_\Sigma x ~, 	\qquad &&\mathrm{d}'_{\Sigma}\eta = \mathrm{d}_\Sigma b ~,\\
 &\mathrm{d}'_{\Sigma}b^+ = \mathrm{d}_\Sigma \eta^+ ~, \qquad &&\mathrm{d}'_{\Sigma}x^+ = \mathrm{d}_\Sigma \eta ~.\\
\end{aligned}
\eeq

We are going to define the $A$-model hierarchy of observables. 
Let us consider the degree $0$ evaluation map $\ev\colon {\cal M}_\Sigma\times T[1]\Sigma \longrightarrow M$ defined as:
\beq\label{partial_evaluation}
 \ev(\x; u,\theta) = x(u) ~.
\eeq 

Since $\F_\Sigma$ is a vector bundle over $\M_\Sigma$ we can extend $\ev$ to a vector bundle morphism $\widehat\ev\colon\F_\Sigma\times T[1]\Sigma\rightarrow T[1]M$ over $\ev$ by asking that for each $f\in C^\infty(M)$ we have 
\[
 \widehat\ev^*\mathrm{d}f = (Q_{\mathrm{BV}} + \mathrm{d}_\Sigma)\ev^*f ~. 
\]
We then compute
\[
 \widehat\ev^* \mathrm{d} x^\mu = (Q_{\mathrm{BV}} + \mathrm{d}_\Sigma) x^\mu = b^\mu + \mathrm{d}_\Sigma x^\mu ~.
\]
For every $\omega\in\Omega^\bullet M$ we can associate a functional $A_\omega$
\beq\label{A_observable}
 A_{\omega}\equiv \widehat\ev^* \omega = \omega(x, b {+}\d_\Sigma x)
\eeq
satisfying by construction $ (Q_{\mathrm{BV}}{+}{\mathrm d}_{\Sigma})A_\omega =A_{\mathrm{d}\omega}$. 
If then $\mathrm{d}\omega=0$ we say that $A_\omega$ is the $A$-model hierarchy of observables of the PSM associated to~$\omega$.

The AKSZ hierarchy described in the previous section, after the identification given between $T[1]M$ and $T^*[1]M$ given by $\alpha$, is defined for each $\omega\in\Omega M$ as 
\begin{equation}\label{AKSZ_observables_PSM}
 {\cal O}_{\omega} = \omega(\x(u,\theta),\b(u,\theta))=\EEv^*\omega
\end{equation}
where the evaluation map $\EEv\colon\F_\Sigma\times T[1]\Sigma\rightarrow T[1]M$ defined in~\eqref{evaluation_map} and given by
\begin{equation}
 \EEv(\x,\b,u,\theta) = (\x(u,\theta),\b(u,\theta))
\end{equation}
is a vector bundle morphism over $\Ev\colon {\cal M}_\Sigma\times T[1]\Sigma \longrightarrow M $ defined as
\beq
 \Ev(\mathbf{x}; u,\theta)=\mathbf{x}(u,\theta)\;.
\eeq
The two morphisms $\ev$ and $\Ev$ are homotopic with homotopy $k\colon{\cal M}_\Sigma\times T[1]\Sigma\times[0,1] \longrightarrow M $ given by
\beq\label{homotopy}
 k(\mathbf{x}; u,\theta;t)= \mathbf{x}(u,t\theta)= x(u) + t\eta^+(u) + t^2 b^+(u) ~.
\eeq
We extend it to the vector bundle morphism $\widehat{k}\colon\F_\Sigma\times T[1]\Sigma\times T[1] I\rightarrow T[1] M$ over~$k$ by imposing that for each~$f\in C^\infty(M)$ we have
\[
 \widehat{k}^*\mathrm{d}f = (Q_{\mathrm{BV}} + \mathrm{d}_\Sigma+ \mathrm{d}_I) k^*f ~,
\]
where $\mathrm{d}_I$ is the de~Rham differential of~$I=[0,1]$. 
We compute
\begin{equation}\label{homotopy1}
 \widehat{k}^* \mathrm{d} x^\mu = \delta X^\mu + t\delta\eta^{+\mu} + t^2\delta b^{+\mu} + (1-t) \mathrm{d}_\Sigma x^\mu + t(1-t)\mathrm{d}_\Sigma\eta^{+\mu} 
  + \mathrm{d}t(\eta^{+\mu} + 2t b^{+\mu}) ~.
\end{equation}
We then define $K(\omega)=\int_{[0,1]} \widehat{k}^*(\omega)$ for each $\omega\in\Omega M$. 
By construction we have that $\widehat{k}^*\omega|_{t=0}=\ev^*\omega=A_\omega$ and $\widehat{k}^*\omega|_{t=1} = \EEv^*\omega={\cal O}_\omega$ and
\beq\label{A_AKSZ_obs_homotopy}
 \mathcal{O}_\omega-A_\omega = K(\d\omega) - (Q_{\mathrm{BV}} {+} \mathrm{d}_\Sigma)K(\omega)~.
\eeq
It is now a direct computation to check that the homotopy~$K$ coincides with the one defined in~\cite{bci:PSMvsA-mod}.

Let us finally discuss the gauge fixing. 
Let us introduce the complex structures $\epsilon$ on~$\Sigma$ and~$J$, compatible with~$\alpha$, on~$M$. 
We denote the holomorphic coordinates as~$z$ and~$x^i$ on~$\Sigma$ and~$M$.
Let us choose the complex gauge fixing for the superfields $\x$ and~$\b$ introduced in~\cite{bonechi:PSM_on_sph.} and discussed in~\cite{bci:PSMvsA-mod} so that we recover the $A$-model action for that sector. 
The gauge fixing lagrangian~$\L$ on the~$A$-model sector is defined by
\begin{equation}\label{A_model_gauge_fixing}
 x^+=b^+=\eta_{z i}=\eta_{\bar z\bar\imath} = \eta^{+ i}_{z} = \eta^{+ \bar\imath}_{\bar z}=0 ~.
\end{equation}

The gauge fixed action reads
\beq\label{A_model_action}
 S_{\mathcal{L}_{\epsilon J}} = \underset{\Sigma}{\int} 
  \Big( -\mathrm{i}p_{z\bar{\jmath}}\partial_{\bar{z}} x^{\bar{\jmath}} +\mathrm{i} p_{\bar{z}i} \partial_z x^i 
   - \mathrm{i}\eta^{+i}_{\bar{z}} D_zb_i + \mathrm{i}\eta^{+\bar{\jmath}}_z D_{\bar{z}}b_{\bar{\jmath}} \\
 + g^{k\bar{r}}R^l_{k\bar{\jmath}i} \eta^{+i}_{\bar{z}} \eta^{+\bar{\jmath}}_z b_l b_{\bar{r}} 
  + g^{i\bar{\jmath}}p_{\bar{z}i} p_{z\bar{\jmath}} \Big)	~,	
\eeq
where $p_{\bar z i}=\eta_{\bar z j}+\Gamma_{ij}^k\eta^{+ j}_{\bar z}b_k$.
Variables appearing in~\eqref{A_model_gauge_fixing} are the momenta of a symplectic tubular neighborhood that determines the BV residual symmetry, as explained in the previous section. 
Contrary to AKSZ observables, the $A$-model observables do not depend on the momenta so that their restriction to~$\L_{\epsilon J}$ is closed under the BV residual symmetry.

\section{Equivariant $A$-model from AKSZ}

We discuss in this section a BV approach to the equivariant version of the $A$-model. 
The geometrical setting consists of a Poisson manifold $(M,\alpha)$ with an action of a Lie group $G$ by Poisson diffeomorphisms. 
We require the existence of an equivariant momentum map $\mu: M\rightarrow \g^*$, where $\g={\rm Lie}~G$.
By momentum map we mean that the fundamental vector fields of the $G$ action are hamiltonian vector field.
We will be mainly interested in the non degenerate case where this is the usual notion of hamiltonian $G$-action.

\subsection{Definition of the model}\label{definition_model}

The model that we are going to discuss was considered in~\cite{zucchini:gaugingPSM,BonechiCabreraZabzine,signori}.
The graded geometric formulation of the equivariant formulation and its AKSZ theory that we are going to use was discussed in~\cite{BonechiCabreraZabzine}.
We briefly recall it. 

The equivariant differential can be described by a hamiltonian vector field~$D$ on the symplectic graded manifold $T^*[1]\big(M\times T[1]\mathfrak{g}[1]\big)$. 
If we take coordinates $(x^\mu,b_\mu)$ on $T^*[1]M$ and $(c^a,\phi^a)$ of degree $(1,2)$ with momenta  $(\xi_a,\widetilde{\xi}_a)$ of degree $(0,-1)$ on $T^*[1]T[1]\g[1]$, we can define the degree~$2$ hamiltonian
\beq\label{eq.ham.}
 \Theta = \frac{1}{2} \alpha^{\mu\nu} b_\mu b_\nu {-} \xi_a\phi^a {-} \mu_a\phi^a + v^\mu_a b_\mu c^a + \frac{1}{2}\xi_a[c,c]^a 
  {+} \widetilde{\xi}_a [c,\phi]^a	~,
\eeq
so that $D(\cdot)=\{\Theta,\cdot\}$ reads:
\beq\label{target_differential}
 &D x^\mu = \alpha^{\mu\nu}b_\nu+ c^a v_a^\mu  ~,\\
 &D b_\mu = \frac{1}{2}\partial_\mu\alpha^{\rho\sigma}b_\rho b_\sigma + \partial_\mu v_a^\rho b_\rho c^a - \phi^a\partial_\mu \mu_a ~,\\
 &D c^a = \phi^a-\frac{1}{2} {f^a_{bd}}c^b c^d  ~,\\
 &D \phi^a = -{f^a_{bc}} c^b \phi^c ~,\\
 &D \xi_a = {v^\mu_a} b_\mu - {f_{ab}^c} \xi_c c^b {- f_{ab}^c} \widetilde{\xi}_c \phi^b  ~,\\
 &D \widetilde{\xi}_a = \xi_a + \mu_a + {f_{ab}^c} \widetilde{\xi}_c c^b  ~.
\eeq
We recover the Kalkman model for Poisson equivariant cohomology as the differential graded subalgebra $W(M,\pi,\mathfrak{g})$ generated by $\{x,b,c,\phi\}$. 
We consider here the case where $\alpha$ is non degenerate and let $b^\mu=\alpha^{\mu\nu}b_\nu$\,.  
We then compute
\beq\label{weyl_model}
 &D x^\mu = b^\mu+ c^a v_a^\mu  ~,\\
 &D b^\mu = -\phi^a v_a^\mu +c^a b^\nu\partial_\nu v_a^\mu  ~.
\eeq
so that $(W(M,\pi,\mathfrak{g}),D)$ coincides with the Kalkman model for equivariant cohomology (see~\cite{kalkman}).

If we look at the target manifold $T^*[1](M\times T[1]\g[1])$ again as a tangent bundle $T[1](M\times \g[1]\times \g^*[-1])$ so that the de~Rham differential is defined as $\d x^\mu=b^\mu$, $\d c^a=\phi^a$ and $\d\widetilde{\xi}_a=\xi_a$ we immediately recognize from (\ref{target_differential},~\ref{weyl_model}) that $D$ can be decomposed~as:
\begin{equation}\label{equivariant_diff_decomposition}
 D= \d+s  ~,\qquad \d s + s \d = s^2 = \d^2 = 0 ~,
\end{equation}
where $s$ is the tangent lift of the target BFV differential giving a resolution for the symplectic quotient.

\begin{remark}\label{tangent_deg}
 The identification of the target manifold with $T[1](M\times\g[1]\times\g^*[-1])$ can be expressed by defining the tangent fibre degree as $\deg x=\deg c=\deg \widetilde{\xi}=0$ and $\deg b=\deg \phi=\deg \xi=1$. 
 Moreover, $\deg \d=1$ and $\deg s=0$.
 Following~\cite{BonechiCabreraZabzine}, the antighost degree $\ag= -\gh+\deg$, where $\gh$ is the natural degree of the target graded manifold, gives the target manifold 
 the structure of $BFV$ manifold, a model for the symplectic reduction of $T^*[1]M$ with respect to the constraints $\mu=0$ and $v_a^\nu b_\nu=0$. 
 We recall that the BFV (Batalin-Fradkin-Vilkovisky) manifolds in general give an homological resolution of constrained system and can be seen as a mathematical formulation of BRST in the hamiltonian setting (see~\cite{BFV, St_constrain}). 
\end{remark}

\begin{remark}\label{cohomology_kalkman}
The map $\varphi:W\rightarrow W$ defined as $\varphi(x,b,c,\phi)=(x,\widetilde{b},c,\widetilde{\phi})$ where
\begin{equation}
 \label{iso_kalk_deRham}\widetilde{\phi}=\phi-\frac{1}{2}[c,c]\,,\;\,\; \widetilde{b}^\mu =b^\mu + c^a v_a^\mu
\end{equation}
intertwines $D|_W$ and the de~Rham differential~$\d$. 
Since the Lie algebra part is acyclic, the cohomology of $(W,D)$ then coincides with $H_{\mathrm{dR}}(M)$. 
Let us introduce the contraction operator $\iota_a=\frac{\partial\ }{\partial c^a}$ and Lie derivative $L_a$ on the Lie algebra variables; then we can 
write the Kalkman differential as
\[
 D|_W = \d_M + c^a(L_{v_a}+L_a) - \phi^a(\iota_{v_a}-\iota_a) ~,
\]
where $\d_M$ denotes the de Rham differential on $M$, $L_{v_a}$ and $\iota_{v_a}$ are the usual Lie derivative and contraction operators on forms, respectively.
We then see that the subcomplex $W'=\bigcap_a \left(\ker\iota_a\cap\ker (L_a+L_{v_a})\right)\subset W$ of elements that are independent on $c$ and $\g$-invariant coincides with the Cartan model for equivariant cohomology.
\end{remark}

Let us now consider the AKSZ sigma model with source $\big(T[1]\Sigma,\mathrm{d}_\Sigma\big)$ and target $T^*[1]\big(M\times T[1]\mathfrak{g}[1]\big)$ with differential~$D$.
We can introduce the superfields:
\beq
 &\x = x + \eta^+ + b^+ ~, \qquad  			&&\b = b + \eta + x^+ ~,\\
 &\c = c + A + \xi^+ ~, \qquad 				&&\superXi = \xi + A^+ + c^+ ~,\\
 &\superPhi = \phi + \psi + \widetilde{\xi}^+ ~, \qquad 	&& \widetilde{\mathbf{\Xi}} = \widetilde{\xi} + \psi^+ + \phi^+ ~.
\eeq
The cohomological BV vector field is $Q_{\mathrm{BV}} = \widetilde{D} - \mathrm{d}'_\Sigma$, where $\widetilde{D}$ is the vector field obtained by composing maps with the (infinitesimal) diffeomorphism of the target defined by~$D$. 
Recalling that $\mathbf{b}^\mu=\alpha^{\mu\nu}(\mathbf{x})\mathbf{b}_\nu$, it acts on the fields $x,b,c,A,\phi,\psi$ as
\beq\label{Weyl_transf}
 &Q_{\mathrm{BV}} x^\mu = b^\mu + c^av_a 	~,\\
 &Q_{\mathrm{BV}} b^\mu = -\partial_\nu v^\mu_a b^\nu c^a - v^\mu_a \phi^a 	~,\\
 &Q_{\mathrm{BV}} c^a = \phi^a - \frac{1}{2}[c,c]^a	~,\\
 &Q_{\mathrm{BV}} A^a = \psi - [c,A]^a - \mathrm{d}_\Sigma c 	~,\\
 &Q_{\mathrm{BV}} \phi^a = - [c,\phi]^a 	~,\\ 
 &Q_{\mathrm{BV}} \psi^a = - [c,\psi]^a - [A,\phi]^a - \mathrm{d}_\Sigma \phi^a ~.
\eeq

We finally write the AKSZ action as
\beq\label{eq.act.}
 S_{\mathrm{BV}} = \int_{T[1]\Sigma} \frac{1}{2}\alpha^{\mu\nu}(\mathbf{x})\mathbf{b}_\mu\mathbf{b}_\nu 
  - \mathbf{\Xi}_a\mathbf{\Phi}^a - \mu_a(\mathbf{x})\mathbf{\Phi}^a + v^\mu_a(\mathbf{x}) \mathbf{b}_\mu \mathbf{c}^a 
   + \frac{1}{2}\mathbf{\Xi}_a[\mathbf{c},\mathbf{c}]^a \\
 + \widetilde{\mathbf{\Xi}}_a[\mathbf{c},\mathbf{\Phi}]^a - \mathbf{b}_\mu\mathrm{d}_\Sigma \mathbf{x}^\mu 
  - \mathbf{\Xi}_a\mathrm{d}_\Sigma \mathbf{c}^a - \widetilde{\mathbf{\Xi}}_a\mathrm{d}_\Sigma \mathbf{\Phi}^a   ~.
\eeq

\subsection{Equivariant $A$-model and AKSZ observables}

We want to define here the analogue of $A$-model observables for the equivariant model.
Let us look for a map analogue to the partial evaluation map defined in~\eqref{partial_evaluation}. 
Since the target space is the shifted tangent bundle $T[1]\M$ with $\M=M\times \g[1]\times \g^*[-1]$ the space $\F_\Sigma$ of AKSZ fields is $T[1]\Map(T[1]\Sigma,\M)$; we then start with a map 
\[
 \mathrm{ev} \colon\mathrm{Map}(T[1]\Sigma,\M)\times T[1]\Sigma\rightarrow \M
\]
defined as
\beq
 \mathrm{ev}(\x,\c,\widetilde{\superXi};u,\theta) = (x(u),c(u)+A(u,\theta),0) ~.
\eeq
Since the target space differential~\eqref{equivariant_diff_decomposition} is not simply the de~Rham differential, on forms we do not take the pull-back of $\ev_0$, as in the previous section, but we look for a vector bundle morphism $\widehat{\ev} \colon\F_\Sigma\times T[1]\Sigma\rightarrow T[1]\M$ over $\ev_0$ that intertwines the differential $Q_{\mathrm{BV}} + \mathrm{d}_{\Sigma}$ with the target differential~$D$,~{\it i.e.} 
\[
 \widehat{\ev}^*D\omega= (Q_{\mathrm{BV}}+\mathrm{d}_\Sigma) \widehat{\ev}^*\omega ~.
\]
From the discussion in Remark~\ref{tangent_deg}, we can conclude that $\widehat{\ev}$ is completely fixed by $\ev$: indeed the equivariant differential decomposes as $D=\d+s$ with $\deg s =0$ so that for each $f\in C(\M)$ we have
\[
 \widehat{\ev}^* \d f = \widehat{\ev}^*(D-s)f = (Q_{\mathrm{BV}}+\d_\Sigma)\ev^*f -\ev^* s f ~.
\]
We then compute
\beq\label{Ev*}
 \widehat{\ev}^* b &= \widehat{\ev}^* \d x =  \widehat{\ev}^* (Dx -c^av_a) = (Q_{\mathrm{BV}}+\mathrm{d}_\Sigma) x - (c^a +A^a)v_a \\
  &= b + \mathrm{d}_\Sigma x - A^a v_a ~, \\
 \widehat{\ev}^* \phi &= \widehat{\ev}^* \d c = \widehat{\ev}^* \Big(Dc + \frac{1}{2}[c,c]\Big) 
  = (Q_{\mathrm{BV}}+\mathrm{d}_\Sigma) (c+A) + \frac{1}{2}[c+A,c+A] \\
 &= \phi + \psi+F(A) ~,\\
 \widehat{\ev}^* \xi_a &= \widehat{\ev}^* \d \widetilde\xi_a = \widehat{\ev}^* ( D\widetilde\xi_a - [\widetilde{\xi},c]_a-\mu_a)= - \mu_a ~.
\eeq
We then finally define for each $\omega(x,c,\widetilde\xi,b,\phi,\xi)\in C(T[1](M\times\g[1]\times\g^*[-1]))$ the following functional
\begin{equation}\label{equivariant_A_observable}
 A_\omega:= \widehat{\ev}^*\omega = \omega(x,c+A,0,b+\mathrm{d}_\Sigma x- A^a v_a,\phi+\psi+F(A), -\mu)\:.
\end{equation}
If $D\omega=0$ then by construction $(Q_{\mathrm{BV}}+\mathrm{d}_\Sigma) A_\omega = 0$ and we say that $A_\omega$ is the $A$-model observable associated to~$\omega$. In particular we will associate to every equivariantly closed form an observable.

Recall that the AKSZ observable associated to~$\omega$ is $\mathcal{O}_\omega = \EEv^*(\omega)$, the pullback of~$\omega$ along the evaluation map $\EEv\colon \F_\Sigma\times T[1]\Sigma\rightarrow T[1]\M$ with
\[
 \EEv(\x,\c,\superXi,\b,\superPhi,\widetilde\superXi;u,\theta)=(\x(u,\theta),\c(u,\theta),\superXi(u,\theta),\b(u,\theta),\superPhi(u,\theta),\widetilde\superXi(u,\theta)) ~.
\]
The map $\EEv$ is a bundle map over $\Ev \colon \Map(T[1]\Sigma,\M)\times T[1]\Sigma\rightarrow \M $ defined as
\[
 \Ev(\x,\c,\widetilde{\superXi};u,\theta)=(\x(u,\theta),\c(u,\theta),\widetilde\superXi(u,\theta)) ~.
\]

We discuss now a homotopy between the $A$-model and AKSZ observables generalizing the discussion that we had in the previous section.
We start with the following homotopy between $\ev_0$ and (the restriction of)~$\Ev$
\[
 \kappa \colon \Map(T[1]\Sigma,\M)\times T[1]\Sigma\times [0,1]\rightarrow \M
\]
defined as
\begin{equation}\label{equivariant_homotopy_0}
 \kappa(\x,\c,\widetilde{\mathbf{\Xi}};u,\theta,t)= (x + t\eta^+ + t^2 x^+,c + A + t^2 \xi^+,t\widetilde\superXi(u,\theta)) ~.
\end{equation}
We then look for 
\[
 \widehat{\kappa} \colon \F_\Sigma\times T[1]\Sigma\times T[1][0,1]\rightarrow T[1]\M
\]
over $\kappa$ so that 
\beq\label{eq.hom.}
 (Q_{\mathrm{BV}}+\d_\Sigma + \d_I) \widehat{\kappa}^* = \widehat{\kappa}^* D~,
\eeq
where $\mathrm{d}_I$ is the de~Rham differential of $I=[0,1]$. 
Again $\widehat\kappa$ is completely determined by $\kappa$ and moreover by construction
\[
 \widehat\kappa^*\omega |_{t=0}= A_\omega\,,\;\;\; \widehat\kappa^*\omega |_{t=1} = {\cal O}_\omega ~.
\]
We then compute 
\beq
 \widehat\kappa^*b^\mu &= \widehat\kappa^* \mathrm{d}x^\mu = (Q_{\mathrm{BV}}+\mathrm{d}_\Sigma+ \mathrm{d}_I) \kappa^*x^\mu - \kappa^*(c^av^\mu_a) \\
 &= b^\mu + (1-t)\mathrm{d}_\Sigma x^\mu - (1-t)A^a v^\mu_a + t(1-t) \partial_\nu v_a^\mu \eta^{+\nu} + t(1-t)\mathrm{d}_\Sigma \eta^{+\mu} + \\
 & + \mathrm{d}t (\eta^{+\mu}+2tb^{+\mu})  + t^2 \delta b^{+\mu} + t \delta \eta^{+\mu} ~,\\
 \widehat\kappa^*\phi &= \widehat\kappa^* \mathrm{d}c = (Q_{\mathrm{BV}}+\mathrm{d}_\Sigma+ \mathrm{d}_I) \kappa^*c + \frac{1}{2}[\kappa^*c,\kappa^*c] \\
 &= \phi+\psi+(1-t^2)F(A) + t^2 \widetilde{\xi}^+ + 2t \xi^+ \mathrm{d}t  ~,	\\
 \widehat\kappa^*\xi &= \widehat\kappa^* \d\widetilde\xi = (Q_{\mathrm{BV}} + \mathrm{d}_\Sigma + \mathrm{d}_I)(t\widetilde\superXi) - t[\widetilde\superXi,\kappa^* c]-\kappa^*\mu ~.
\eeq
If we define $\mathcal{K}(\cdot)=\int_I \widehat{\kappa}^* (\cdot)$\,, we get 
\beq\label{A_AKSZ_homotopy}
 \mathcal{O}_\omega- A_\omega = \mathcal{K} (D\omega) - (Q_{\mathrm{BV}}+\mathrm{d}_\Sigma)\mathcal{K} \omega	~.
\eeq

Of course, if we set to zero all the variables associated to the Lie algebra $\mathfrak{g}$\,, we recover the homotopy between the AKSZ observables of the PSM and the $A$-model observables described in~\eqref{homotopy1}.

\subsection{Cohomology of $D$}

An interesting consequence of~\eqref{A_AKSZ_homotopy} is the following characterization of the cohomology of the target differential $D$ defined in~\eqref{target_differential}.
Indeed, after the restriction to zero form observables, $\d_\Sigma$ does not appear in~\eqref{A_AKSZ_homotopy} and $Q_{BV}$ acts as $D'$, the vector field obtained composing the maps of $\F_\Sigma$ with the infinitesimal diffeomorphism of the target defined by~$D$. 
In other terms it does not involve derivatives with respect to the source coordinates, so that it is a pointwise relation that can be read as a relation defined on the target as follows.

Let $i\colon W(M,\pi,\g)\rightarrow {\cal A}\equiv(C(T^*[1](M\times T[1]\g[1])),D)$ be the injection of the Kalkman model described in Subsection~\ref{definition_model} and let $p:{\cal A}\rightarrow W(M,\pi,\g)$ be the quotient map with kernel generated by $\xi+\mu$ and~$\widetilde\xi$. 
It is a direct check to verify that $p$ is a chain map. 
Clearly we have that $p\circ i = {\rm id}_W$. 
Now it is clear that~\eqref{A_AKSZ_homotopy} for forms of degree $0$ translates into 
\begin{equation}\label{target_homotopy}
 {\rm id}_{\cal A}- i\circ p = \mathcal{K}_0\circ D-D\circ\mathcal{K}_0
\end{equation}
where 
\[
 \mathcal{K}_0(\omega)=\int_I \omega(x,c,t\widetilde\xi,b,\phi,t\xi -(1-t)\mu+ \d_I t\ \widetilde\xi)
\]
for each $\omega\in{\cal A}$. 
We can then conclude that $i$ and $p$ are inverse up to homotopy so that the cohomology of $D$ is isomorphic to the cohomology of the Kalkman complex 
(or equivalently of the Weil complex, see~\cite{kalkman}) that is de~Rham cohomology $H_{\mathrm{dR}}(M)$ (see Remark~\ref{cohomology_kalkman}). 
It is maybe useful to stress that we are not restricting it to the subcomplex $W'$ giving equivariant cohomology. 
Finally, if $\varphi \colon W\rightarrow W$ is the isomorphism defined in~\eqref{iso_kalk_deRham} and $\Theta$ is the degree $2$ hamiltonian in~\eqref{eq.ham.}, then $\varphi(p(\Theta))=-\alpha$, so that we can say that $-\Theta$ represents in $\cal A$ the class of the symplectic form in the de~Rham cohomology of~$M$.

\section{Gauge fixing}

We discuss here two different gauge fixings of the AKSZ theory discussed in the previous section. 
The $A$-model sector is always gauge fixed with the complex gauge fixing defined in~\eqref{A_model_gauge_fixing}.

In both cases the Lagrangian gauge fixing $\L$ is given together with an adapted symplectic tubular neighborhood, {\it i.e.} a symplectomorphism between the BV space of fields 
$\F_\Sigma$ and $T^*[-1]\L$ that fixes also a residual BV symmetry as explained in Section~\ref{AKSZ_background}. 
The $A$-model observables do not depend on the momenta so that after the restriction they are invariant under the residual BV symmetry.

\subsection{Gauge fixing for the symplectic reduction}

We assume that $\partial_k v^{\bar\imath}_a=0$, {\it i.e.} the real $G$~action on~$M$ gives rise to an holomorphic action of $G_{\mathbb C}$. 
It can be checked that, once we assume the complex gauge fixing~\eqref{A_model_gauge_fixing}, the ghost~$c$ disappears from the action.

According to the discussion in Remark~\ref{tangent_deg}, the target manifold of the AKSZ construction is a BFV~space, {\it i.e.} a model for the symplectic reduction of $T^*[1]M$ with respect to the graded constraints $\mu=0$ and~$v^\nu b_\nu=0$. 
If the $G$~action is free on~$\mu=0$ then $\mu^{-1}(0)/G$ is smooth and the reduced space is $T^*[1](\mu^{-1}(0)/G)$. 
In this case the BV theory should be regarded as equivalent to the Poisson Sigma Model with the reduced target. 
According to the discussion in~\cite{BonechiCabreraZabzine}, the natural gauge fixing of the Lie algebra sector is defined by putting the antighosts variables to zero; this means:
\begin{equation}\label{Lie_algebra_gauge_fixing}
 \mathbf{\Xi}=\widetilde{\mathbf{\Xi}}=0 ~.
\end{equation}

The residual gauge symmetry $Q_\L$ given by this symplectic tubular neighbourhood is directly read from~\eqref{Weyl_transf} together with
\beq\label{weil_transf2}
 Q_\L \eta_{\bar{z} i} &= \partial_i \alpha^{k\bar{\jmath}} \eta_{\bar{z} k} b_{\bar{\jmath}} 
  + \partial_i \mu_a \psi^a_{\bar z} + \partial_i\partial_j \mu_a \eta^{+ j}_{\bar z} \phi^a+\\
  &\quad + \partial_i v_a^j (\eta_{\bar{z} j} c^a + b_j\psi^a_{\bar z} ) + \partial_i\partial_j v^k_a \eta^{+j}_{\bar z} b_k c^a ~,\\
 Q_\L \eta^{+ i}_{\bar z} &= - \partial_z x^i + \partial_k \alpha^{i\bar{\jmath}}\eta^{+k}_{\bar{z}} b_{\bar \jmath} 
  + v^i_a A^a_{\bar z} +\partial_k v^i_a \eta^{+ k }_{\bar z} c^a ~,\\
 Q_{\L} \xi^+ &= \widetilde{\xi}^+ - F(A) - [c,\xi^+]\;,\\
 Q_{\L}\widetilde{\xi}^+ &= - \mathrm{d}_\Sigma\psi -[A,\psi] - [c,\widetilde{\xi}^+]-[\xi^+,\phi] ~.
\eeq

Since the $A$-model observables defined in~\eqref{equivariant_A_observable} are independent on the coordinates~\eqref{A_model_gauge_fixing} and~\eqref{Lie_algebra_gauge_fixing}, they are also invariant when restricted to the gauge fixing lagrangian under~$Q_\L$. 
This is not true for the AKSZ observables.

After the introduction of an arbitrary affine connection $\Gamma$ and the definition of $p_{\bar z j}=\eta_{\bar z j}+\Gamma_{ji}^k\eta^{+ i}_{\bar z}b_k$, we obtain the gauge-fixed action $\mathcal{S}_\mathcal{L} = \mathcal{S}_{\mathcal{L}_{\epsilon J}} + \mathcal{S}_{\mathcal{L}_{\mathfrak{g}}}$  where
$\mathcal{S}_{\mathcal{L}_{\epsilon J}}$ is computed in~\eqref{A_model_action}~and
\beq
 S_{\mathcal{L}_{\mathfrak{g}}} = \int_\Sigma \Big(&\mu_a \widetilde{\xi}^{+ a} 
  + \partial_i \mu_a \eta^{+ i}_{\bar z}\psi^a_z + \partial_{\bar \imath}\mu_a \eta^{\bar\imath}_z \psi^a_{\bar z} 
   + \partial_i\partial_{\bar\jmath} \mu_a \eta^{+ i}_{\bar z} \eta^{+ \bar\jmath}_z\phi^a +v_a^i b_i \xi^{+ a} + v_a^i p_{i\bar z} A^a_z + \\
 &+ \nabla_k v_a^i \eta^{+ k}_{\bar z} b_i A^a_z 
  + v_a^{\bar\imath} b_{\bar \imath} \xi^{+ a} + v_a^{\imath} p_{\bar\imath z} A^a_{\bar z} 
   + \nabla_{\bar k} v_a^{\imath} \eta^{+ \bar k}_{z} b_{\bar\imath} A^a_{\bar z} \Big) ~.
\eeq

This action is quadratic in the fields $p$, which can then be integrated out.
Their equations of motion are $p_{\bar{z}i} = \alpha_{i\bar{\jmath}} \partial_{\bar{z}}x^{\bar{\jmath}} - \alpha_{i\bar{\jmath}} v_a^{\bar{\jmath}} A_{\bar{z}}^a$ and the effective action obtained with this integration is thus:
\beq
 S_{\L}= \int_\Sigma  \Big( &\eta^{+i}_{\bar{z}} D_zb_i
  +\eta^{+\bar{\jmath}}_z D_{\bar{z}}b_{\bar{\jmath}} + \alpha^{\bar{r}k}R^l_{k\bar{\jmath}i} \eta^{+i}_{\bar{z}} \eta^{+\bar{\jmath}}_z b_l b_{\bar{r}} 
   + \alpha_{i\bar{\jmath} }\partial_A x^i \bar\partial_Ax^{\bar{\jmath}} + \\
 &+ \nabla_l v_a^k \eta^{+l}_{\bar{z}}b_k A_z^a 
  + \nabla_{\bar \jmath} v_a^{\bar k} \eta^{+\bar \jmath}_{z}b_{\bar k} A_{\bar z}^a
   +\big( v_a^i b_i + v_a^{\bar{\jmath}}b_{\bar{\jmath}}\big)\xi^{+a}_{z\bar{z}}  + \\
 & + \mu_a\widetilde{\xi}^{+a} + \partial_{\bar{\jmath}}\mu_a \eta^{+\bar{\jmath}}_{z}\psi_{\bar{z}}^a 
  + \partial_i\mu_a \eta^{+i}_{\bar{z}}\psi_z^a
   + \partial_i\partial_{\bar{\jmath}} \mu_a \eta^{+\bar{\jmath}}_{z}\eta^{+i}_{\bar{z}} \phi^a \Big) ~,
\eeq
where $\partial_A x^i= \partial_z x^i + v_a^i A^a_z$. 
The dependence on the connection $A$ is now at most quadratic and the quadratic term is non degenerate if $\det(v_a,v_b)\not=0$. 
This is guaranteed if the $G$ action is free on $\mu^{-1}(0)$, so that this action is well defined when the symplectic reduction $\mu^{-1}(0)/G$ is smooth.

By construction the Lie algebra fields are Lagrange multipliers that constrain the system to $\mu_a(X+\eta^+)=0$ and $v_a(x+\eta^+)^i(b+\eta)_i=0$. 
We know that in K\"ahler reduction $T(\mu^{-1}(0))/G$ is realized as the subbundle $J(\g_M)^\perp\cap\g_M^\perp\subset T(\mu^{-1}(0))$, where $\g_M$ denotes the bundle spanned by the $\g$-vectors. 
The zero and one form component of $\mu(x+\eta^+)=0$ force then the field $x$ to take values in $\mu^{-1}(0)$ and $\eta^+$ in $T(\mu^{-1}(0))/G$.
The two form constraint $\partial_i\partial_{\bar\jmath}\mu_a\eta^{+i}_{\bar z}\eta^{+\bar\jmath}_z=0$ is instead just a consequence of the interplay between the constraint on superfields and the complex gauge fixing and it has not a geometric origin.  
One way to avoid it is to modify~\eqref{Lie_algebra_gauge_fixing} to
\begin{equation}\label{Lie_algebra_gauge_fixing_2}
 c=0~, \quad \superXi = c^+ \,, \quad \widetilde\superXi=0 ~.
\end{equation}
This fixes the ghost~$c$ and makes~$c^+$ be the multiplier for~$\phi=0$ so that the above undesired constraint disappears. 
It must be clear that with this choice the $A$-model observables depending on $c$ are not anymore invariant under the residual gauge fixing.

\subsection{The gauge multiplet and topological Yang-Mills}

We consider here a different gauge fixing of the Lie algebra sector that recovers the so called topological Yang-Mills theory in two dimensions, considered by Witten in~\cite{Witten:1992xu}.  
This connection was already established in~\cite{zucchini:gaugingPSM}; here we use a slightly different gauge fixing and emphasize the relation between the residual gauge symmetry and the gauge multiplet of supersymmetry. 
The basic tool for introducing topological Yang-Mills theory is the gauge multiplet of 2d supersymmetry. 
In our BV framework it must appear as a residual BV symmetry of the gauge fixed action.
We have first to recognize all the fields needed to reconstruct the gauge multiplet. 
The gauge multiplet consists in the following fields 
\beq
\begin{tabular}{l*{7}{c}}
 		&$\phi$	&$\psi$	&$A$	&$H$	&$\chi$			&$\eta$	&$\lambda$ 	\\
 \hline
 ghost \#	&2	&1	&0	&0	&-1			&-1	&-2		\\
 \hline
 form \#	&0	&1	&1	&0	&0			&0	&0		\\
\end{tabular}
\eeq
where the parity is the ghost modulo~$2$. 
It is easy to see that we already have almost all these fields by doing the following matches
\beq
\begin{tabular}{l*{7}{c}}
 AKSZ		&$\phi$	&$\psi$	&$A$	&$\xi$	&$\widetilde{\xi}$	&-	&-		\\
 \hline
 gauge multiplet		&$\phi$	&$\psi$	&$A$	&$H$	&$\chi$			&$\eta$	&$\lambda$ 	\\
\end{tabular}
\eeq
The fields~$\eta$ and~$\lambda$ do not appear in the PSM but can be introduced as a trivial pair.
Let us define the trivial pair $\lambda,\rho\in\Omega^0(\Sigma;\mathfrak{g})$ of ghost number~$-2$ and~$-1$ respectively and with momenta $\lambda^+,\rho^+\in\Omega^2(\Sigma;\mathfrak{g}^*)$ of ghost degree~$1$ and~$0$\,.
The BV action will be shifted to
\beq
 S_{\mathrm{BV}}' := S_{\mathrm{BV}} + \int_\Sigma \lambda^+_a \rho^a
\eeq
and correspondingly we have the following action of the BV symmetry for~$\lambda$ and~$\rho$\,:
\beq
 Q_{\mathrm{BV}} \lambda^a = \rho^a ~,	\qquad  Q_{\mathrm{BV}} \rho^a = 0 ~.
\eeq
If we define $\zeta:= \rho + [c,\lambda]$ we get the transformations:
\beq
 Q_{\mathrm{BV}} \lambda = \zeta - [c,\lambda] ~,	\qquad  Q_{\mathrm{BV}} \zeta = [\phi,\lambda] - [c,\zeta] ~.
\eeq
The gauge multiplet is reconstructed with $\zeta\sim \eta$.

We can now collect the action of $Q_{\mathrm{BV}}$ on these fields
\beq\label{BV_supersymmetry}
 &Q_{\mathrm{BV}} \phi= -[c,\phi] ~,		    	    	\qquad 	&&Q_{\mathrm{BV}} \psi= -[c,\psi] - [A,\phi] - \mathrm{d}_\Sigma \phi ~,	\\
 &Q_{\mathrm{BV}} A= \psi - [c,A] -\mathrm{d}_\Sigma c~,	    	\qquad 	&&Q_{\mathrm{BV}} \xi= - [\xi,c] - [\widetilde{\xi},\phi] ~,\\
 &Q_{\mathrm{BV}} \widetilde{\xi}= \xi - [\widetilde{\xi},c] ~, 	\qquad  &&Q_{\mathrm{BV}} \lambda = \zeta - [c,\lambda] ~,\\
 &Q_{\mathrm{BV}} \zeta = [\phi,\lambda] - [c,\zeta] 	\qquad 	&& Q_{\mathrm{BV}} c = \phi - \frac{1}{2}[c,c]~.		
\eeq
We then see that 
\[
 Q_{\mathrm{BV}} = \delta_{\mathrm{BRST}} + \delta_{\mathrm{susy}} ~,
\]
{\it i.e.} it encodes the superymmetry and the BRST transformation of the gauge multiplet. 

The action of topological Yang-Mills is recovered by defining the Lagrangian~$\L_f$ with the gauge-fixing fermion~$f$ defined as
\beq
 f= \int_\Sigma \frac{1}{2} \langle \widetilde{\xi},\star\xi \rangle + \langle D_A\lambda ,\psi \rangle ~,
\eeq
where $\star$ is the Hodge star for a metric on~$\Sigma$ and $\langle,\rangle$ is a non degenerate invariant bilinear form on~$\g$. 
\beq
 S_{\mathcal{L}_f} &= S_{\mathcal{L}_0} + Q_{\mathrm{BV}} f = Q_{\mathrm{BV}} \Big(f - \int_\Sigma \langle \widetilde{\xi}, F_A\rangle\Big)\\
  &= \int_\Sigma \Big(\frac{1}{2} \langle \xi , \star \xi \rangle - \langle \xi , F_A \rangle - \langle \widetilde{\xi} , D_A \psi \rangle
   + \langle D_A\zeta, \psi \rangle  + \\
 &\qquad\qquad+ \langle D_A \lambda , D_A \phi \rangle + \frac{1}{2}\langle [\widetilde{\xi},\widetilde{\xi}], \star\phi \rangle 
  + \langle [\psi,\lambda], \psi \rangle	\Big)~.
\eeq

The residual BV symmetry is then given by the same formulas as in~\eqref{BV_supersymmetry}. 
In particular, the $A$-model observables are just functions of $(c,\phi,A,\psi)$ and so are invariant. 
The full gauge fixed action is then recovered as:
\beq
 S_{\L}= \underset{\Sigma}{\int} \Big( &\eta^{+i}_{\bar{z}} D_zb_i
  +\eta^{+\bar{\jmath}}_z D_{\bar{z}}b_{\bar{\jmath}} + \alpha^{\bar{r}k}R^l_{k\bar{\jmath}i} \eta^{+i}_{\bar{z}} \eta^{+\bar{\jmath}}_z b_l b_{\bar{r}} 
   + \alpha_{i\bar{\jmath} }\partial_A x^i \bar\partial_Ax^{\bar{\jmath}} + \\
 & + \nabla_l v_a^k \eta^{+l}_{\bar{z}}b_k A_z^a - 
  \nabla_{\bar \jmath} v_a^{\bar k} \eta^{+\bar \jmath}_{{z}}b_{\bar k} A_{\bar z}^a 
   + \partial_{\bar{\jmath}}\mu_a \eta^{+\bar{\jmath}}_{z}\psi_{\bar{z}}^a  +\\
 & + \partial_i\mu_a \eta^{+i}_{\bar{z}}\psi_z^a + \partial_i\partial_{\bar{\jmath}} \mu_a \eta^{+\bar{\jmath}}_{z}\eta^{+i}_{\bar{z}} \phi^a +\\
 & + \frac{1}{2} \langle \xi , \star \xi \rangle - \langle \xi , F_A \rangle - \langle \widetilde{\xi} , D_A \psi \rangle
   + \langle D_A\zeta, \psi \rangle  + \\
 & + \langle D_A \lambda , D_A \phi \rangle + \frac{1}{2}\langle [\widetilde{\xi},\widetilde{\xi}], \star\phi \rangle 
  + \langle [\psi,\lambda], \psi \rangle	\Big)~.
\eeq
Up to residual BV transformations this action depends only on 
the cohomology class of $(\alpha +\mu_a\phi^a)$. 
By construction the BV action~\eqref{eq.act.} is defined by the target data through the AKSZ observable associated to the hamiltonian $\Theta$ in~\eqref{eq.ham.}.  
As we proved, this observable is connected to the corresponding $A$-model observable via the homotopy~$\mathcal{K}$.
Thus the BV action can be decomposed as: kinetic term + $A$-model observable + BV exact term. 
The $A$-model observables are well-behaved under our gauge-fixing, however BV exact terms become generically Q-exact terms only on-shell.  
However if we choose Q-symmetry to close off-shell then the gauge fixed action can be written as A-model observable associated to $(\alpha +\mu_a\phi^a)$ plus Q-exact terms, similar to the construction presented in~\cite{Nekrasov}.


\bibliography{Bibliography}

\end{document}